\documentclass[aps,prl,showpacs,twocolumn,superscriptaddress]{revtex4}
\usepackage{amsmath}
\usepackage{amssymb}
\usepackage{epsfig}
\usepackage{color}
\usepackage{subfigure}
\usepackage{graphicx,amsmath}

\begin{document}
\title{Discovering Strongly Correlated Quantum Spin Liquid}
\author{V. R. Shaginyan}\email{vrshag@thd.pnpi.spb.ru}
\affiliation{Petersburg Nuclear Physics Institute, Gatchina,
188300, Russia}\author{K. G. Popov}\affiliation{Komi Science
Center, Ural Division, RAS, Syktyvkar, 167982, Russia}\author{V. A.
Khodel} \affiliation{Russian Research Centre Kurchatov Institute,
Moscow, 123182, Russia} \affiliation{McDonnell Center for the Space
Sciences \& Department of Physics, Washington University, St.
Louis, MO 63130, USA}

\begin{abstract}
Strongly correlated Fermi systems are among the most intriguing and
fundamental systems in physics. We show that the herbertsmithite
$\rm ZnCu_3(OH)_6Cl_2$ can be viewed as a new type of strongly
correlated electrical insulator that possesses properties of
heavy-fermion metals with one exception: it resists the flow of
electric charge. We demonstrate that herbertsmithite's low
temperature properties are defined by a strongly correlated quantum
spin liquid made with such hypothetic particles as fermionic
spinons which carry spin $1/2$ and no charge. Our calculations of
its thermodynamic and relaxation properties are in good agreement
with recent experimental facts and allow us to reveal their scaling
behavior which strongly resembles that observed in heavy-fermion
metals. Analysis of the dynamic magnetic susceptibility of strongly
correlated Fermi systems suggests that there exist at least two
types of its scaling.
\end{abstract}


\maketitle

Strongly correlated Fermi systems represented by heavy-fermion (HF)
metals are well experimentally studied but only recently have got
an adequate theoretical description \cite{pr}. Landau Fermi liquid
(LFL) theory is highly successful in the condensed matter physics.
The key point of this theory is the existence of fermionic
quasiparticles defining the thermodynamic, relaxation and dynamic
properties of conventional metals. However, strongly correlated
Fermi systems encompassing a variety of systems that display
behavior not easily understood within the Fermi liquid theory and
called non-Fermi liquid (NFL) behavior. A paradigmatic example of
the NFL behavior is demonstrated by HF metals, where a quantum
phase transition (QPT) induces a transition between LFL and NFL
\cite{loh,pr}. QPT can be tuned by different parameters, such as
the chemical composition, the pressure, and the magnetic field.
Magnetic materials, in particular insulators, are interesting
subjects of study due to a quantum spin liquid (QSL) that may
develop in them, defining their low-temperature properties. Exotic
QSL is formed with such hypothetic particles as fermionic spinons
carrying spin $1/2$ and no charge. A search for the materials is a
challenge for condensed matter physics \cite{bal}. In zero and high
magnetic fields $B$
\cite{herb0,herb4,herb1,herb2,mil,herb3,herb,sl,sl2,sl1}, the
experimental studies of herbertsmithite $\rm ZnCu_3(OH)_6Cl_2$ have
discovered gapless excitations, analogous to quasiparticle
excitations near the Fermi surface in HF metals, indicating that
$\rm ZnCu_3(OH)_6Cl_2$ is the promising system to investigate its
QPTs and QSLs \cite{prbr,eplh,pla12}. The observed behavior of the
thermodynamic properties of $\rm ZnCu_3(OH)_6Cl_2$ strongly
resembles that in HF metals since a simple kagome lattice has a
dispersionless topologically protected branch of the spectrum with
zero excitation energy \cite{prbr,green,vol}. This indicates that
QSL formed by the ideal kagome lattice and located near the fermion
condensation quantum phase transition (FCQPT) can be considered as
a strongly correlated quantum spin liquid (SCQSL). This observation
allows us to establish a close connection between $\rm
ZnCu_3(OH)_6Cl_2$ with its SCQSL and HF metals whose HF system is
located near FCQPT and, therefore, exhibiting an universal scaling
behavior \cite{pr,prbr,eplh}. Thus, FCQPT represents QPT of $\rm
ZnCu_3(OH)_6Cl_2$ and both herbertsmithite and HF metals can be
treated in the same framework, while SCQSL is composed of spinons
and these with zero charge and spin $\sigma=\pm1/2$ occupy the
corresponding Fermi sphere with the Fermi momentum $p_F$
\cite{pr,prbr,eplh,pla12}.

In our paper we show that both non-Fermi liquid and scaling
behavior of such strongly correlated Fermi systems as HF metals and
$\rm ZnCu_3(OH)_6Cl_2$ can be described within the frame of the
theory of FCQPT. Analyzing experimental data obtained in
measurements on strongly correlated Fermi systems with different
microscopic properties we have found out that they demonstrate the
universal non-Fermi liquid behavior. Our analysis of the dynamic
magnetic susceptibility of strongly correlated Fermi systems
suggests that there exist at least two types of its scaling. We
calculate the thermodynamic and relaxation properties of
herbertsmithite and HF metals. The calculations are in a good
agreement with experimental data and allow us to detect the
low-temperature behavior of $\rm ZnCu_3(OH)_6Cl_2$ defined by SCQSL
as that observed in heavy fermion metals.

To study theoretically the low temperature thermodynamic,
relaxation and scaling properties of herbertsmithite, we use the
model of homogeneous HF liquid \cite{pr}. This model permits to
avoid complications associated with the crystalline anisotropy of
solids. Similar to the electronic liquid of HF metals, SCQSL is
composed of chargeless fermions (spinons) with $S=1/2$ occupying
the corresponding two Fermi spheres with the Fermi momentum $p_F$.
The ground state energy $E(n)$ is given by the Landau functional
depending on the quasiparticle distribution function $n_\sigma({\bf
p})$, where $p$ is the momentum and $\sigma$ is the spin index. The
effective mass $M^*$ is governed by the Landau equation
\cite{land,pr}
\begin{eqnarray}
\nonumber
&&\frac{1}{M^*(B,T)}=\frac{1}{M^*}+\frac{1}{p_F^2}\sum_{\sigma_1}
\int\frac{{\bf p}_F{\bf p_1}}{p_F}\\
&\times&F_{\sigma,\sigma_1}({\bf p_F},{\bf
p}_1)\frac{\partial\delta n_{\sigma_1}({\bf p}_1,B,T)}
{\partial{p}_1}\frac{d{\bf p}_1}{(2\pi) ^3}. \label{HC3}
\end{eqnarray}
Here we rewrite the quasiparticle distribution function as
$n_{\sigma}({\bf p},B,T) \equiv n_{\sigma}({\bf p},B=0,T=0)+\delta
n_{\sigma}({\bf p},B,T)$. The Landau amplitude $F$ is completely
defined by the fact that the system has to be at FCQPT
\cite{pr,ckz,epl,khodb}, see \cite{ckz,khodb,epl} for details of
solving Eq. \eqref{HC3}. The sole role of Landau amplitude is to
bring the system to FCQPT point, where Fermi surface alters its
topology so that the effective mass acquires temperature and field
dependencies. At this point, the term $1/M^*$ vanishes, Eq.
\eqref{HC3} becomes homogeneous and can be solved analytically
\cite{pr,ckz}. At $B=0$, the effective mass, being strongly $T$ -
dependent, demonstrates the NFL behavior given by Eq. \eqref{HC3}
\begin{equation}
M^*(T)\simeq a_TT^{-2/3}\label{MTT},
\end{equation}
where $a_T$ is a constant. At finite $T$, under the application of
magnetic field $B$ the two Fermi spheres due to the Zeeman
splitting are displaced by opposite amounts, the final chemical
potential $\mu$ remaining the same within corrections of order
$B^2$. As a result, field $B$ drives the system to LFL region, and
again it follows from  Eq. \eqref{HC3} that
\begin{equation}
M^*(B)\simeq a_BB^{-2/3}\label{MBB},
\end{equation}
where $a_B$ is a constant. It is seen from Eqs. \eqref{MTT} and
\eqref{MBB} that effective mass diverges at FCQPT. At finite $B$
and $T$, the solutions of Eq. \eqref{HC3} $M^*(B,T)$ can be well
approximated by a simple universal interpolating function. This
interpolation occurs between the LFL regime, given by Eq.
\eqref{MBB} and NFL regime given by Eq. \eqref{MTT} \cite{pr,ckz}.
Experimental facts and calculations show that $M^*(B,T)$ as a
function of $T$ at fixed $B$ reaches its maximum value $M^*_M$ at
$T_M$, see e.g. \cite{pr}. To study the universal scaling behavior
of strongly correlated Fermi system, it is convenient to introduce
the normalized effective mass $M^*_N$ and the normalized
temperature $T_N$ dividing the effective mass $M^*$ and temperature
$T$ by their values at the maximum, $M^*_M$ and $T_M$ respectively.
In the same way, we can normalize other thermodynamic functions
such as the spin susceptibility $\chi$ and the heat capacity $C$.
As a result, we obtain
\begin{equation}
\chi_N\simeq (C/T)_N\simeq M^*_N, \label{SCB}
\end{equation}
where $\chi_N$ and $(C/T)_N$ are the normalized values of $\chi$
and $C/T$, respectively. We note that our calculations of $M^*_N$
based on Eq. \eqref{HC3} do not contain any fitting parameters. The
normalized effective mass $M^*_N=M^*/M^*_M$ as a function of the
normalized temperature $y=T_N=T/T_M$ is given by the interpolating
function \cite{pr}
\begin{equation}M^*_N(y)\approx c_0\frac{1+c_1y^2}{1+c_2y^{8/3}}.
\label{UN2}
\end{equation}
Here $c_0=(1+c_2)/(1+c_1)$, $c_1$ and $c_2$ are fitting parameters.
Since magnetic field $B$ enters Eq. \eqref{HC3} only in combination
$B\mu_B/k_BT$, we have $T_{\rm max}\propto B$ \cite{ckz,pr}, where
$\mu_B$ is the Bohr magneton and $k_B$ is the Boltzmann constant.
Thus, for finite magnetic fields variable $y$ becomes
\begin{equation}\label{YTB}
y=T/T_{N}\propto k_BT/\mu_BB.
\end{equation}
Since the variables $T$ and $B$ enter symmetrically Eq. \eqref{UN2}
is valid for $y=\mu_BB/k_BT$.

To construct the dynamic spin susceptibility $\chi({\bf
q},\omega,T)=\chi{'}({\bf q},\omega,T)+i\chi{''}({\bf q},\omega,T)$
as a function of momentum $q$, frequency $\omega$ and temperature
$T$, again we use the model of homogeneous HF liquid located near
FCQPT. To deal with the dynamic properties of Fermi systems, one
can use the transport equation describing a slowly varying
disturbance $\delta n_{\sigma}({\bf q},\omega)$ of the
quasiparticle distribution function $n_0({\bf p})$, and $n=\delta
n+n_0$. We consider the case when the disturbance is induced by the
application of external magnetic field $B=B_0+\lambda B_1({\bf
q},\omega)$ with $B_0$ being a static field and $\lambda B_1$ a
$\omega$-dependent field with $\lambda\to0$. As long as the
transferred energy $\omega<qp_F/M^*<<\mu$, where $M^*$ is the
effective mass and $\mu$ is the chemical potential, the
quasiparticle distribution function $n({\bf q},\omega)$ satisfies
the transport equation \cite{PinNoz}
\begin{eqnarray}\label{TREQ}
&&({\bf qv_p}-\omega)\delta n_{\sigma}-{\bf qv_p}\frac{\partial
n_0}{\partial\varepsilon_p}\sum_{\sigma_1{\bf
p}_{1}}f_{\sigma,\sigma_1}({\bf pp}_{1})\delta n_{\sigma_1}({\bf
p}_{1})\nonumber\\&=&{\bf qv_p}\frac{\partial
n_0}{\partial\varepsilon_p}\sigma\mu_B(B_0+\lambda B_1).
\end{eqnarray}
Here $\mu_B$ is the Bohr magneton and $\varepsilon_p$ is the
single-particle spectrum. In the field $B_0$, the two Fermi
surfaces are displaced by opposite amounts, $\pm B_0\mu_B$, and the
magnetization $\mathcal{M}=\mu_B(\delta n_+-\delta n_-)$, where the
two spin orientations with respect to the magnetic field are
denoted by $\pm$, and $\delta n_{\pm}=\sum_p \delta n_{\pm}({\bf
p})$. The spin susceptibility $\chi$ is given by
$\chi=\partial\mathcal{M}/\partial B_{|_{B=B_0}}$. In fact, the
transport equation \eqref{TREQ} is reduced to two equations which
can be solved for each direction $\pm$ and allows one to calculate
$\delta n_{\pm}$ and the magnetization. The response to the
application of $\lambda B_1({\bf q},\omega)$ can be found by
expanding the solution of Eq. \eqref{TREQ} in a power series with
respect to $M^*\omega/qp_F$. As a result, we obtain the imaginary
part of the spin susceptibility
\begin{equation}\label{chi2}
\chi{''}({\bf q},\omega)=\mu_B^2\frac{\omega(M^*)^2}{2\pi
q}\frac{1}{(1+F^a_0)^2},
\end{equation}
where $F^a_0$ is the dimensionless spin antisymmetric quasiparticle
interaction \cite{PinNoz}. The interaction $F^a_0$ is found to
saturate at $F^a_0\simeq -0.8$ \cite{pfw,vollh1} so that the factor
$(1+F^a_0)$ in Eq. \eqref{chi2} is finite and positive. It is seen
from Eq. \eqref{chi2} that the second term is an odd function of
$\omega$. Therefore, it does not contribute to the real part
$\chi'$ and forms the imaginary part $\chi''$. Taking into account
that at relatively high frequencies $\omega\geq qp_F/M^*\ll\mu$ in
the hydrodynamic approximation $\chi'\propto 1/\omega^2$
\cite{forst}, we conclude that the equation
\begin{equation}\label{chi3}
\chi({\bf
q},\omega)=\frac{\mu_B^2}{\pi^2(1+F^a_0)}\frac{M^*p_F}{1+i\pi
\frac{M^*\omega}{qp_F(1+F^a_0)}},
\end{equation}
produces the simple approximation for the susceptibility $\chi$ and
satisfies the Kramers-Kronig relation connecting the real and
imaginary parts of $\chi$.

To understand how can $\chi''$ and $\chi$ given by Eqs.
\eqref{chi2} and \eqref{chi3}, respectively, depend on temperature
$T$ and magnetic field $B$, we recall that near FCQPT point the
effective mass $M^*$ depends on $T$ and $B$. To elucidate a scaling
behavior of $\chi$, we employ Eq. \eqref{MTT} to describe the
temperature dependence of $\chi$. It follows from Eqs. \eqref{chi3}
and \eqref{MTT} that
\begin{equation}\label{SCHI}
T^{2/3}\chi(T,\omega)\simeq \frac{a_1}{1+ia_2E}.
\end{equation}
Here $a_1$ and $a_2$ are constants absorbing irrelevant values and
$E=\omega/(k_BT)^{2/3}$. As a result, the imaginary part
$\chi''(T,\omega)$ satisfies the equation
\begin{equation}\label{SCHII}
T^{2/3}\chi''(T,\omega)\simeq\frac{a_3E}{1+a_4E^2},
\end{equation}
where $a_3$ and $a_4$ are constants. It is seen from Eq.
\eqref{SCHII} that $T^{2/3}\chi''(T,\omega)$ has a maximum
$(T^{2/3}\chi''(T,\omega))_{\rm max}$ at some $E_{\rm max}$ and
depends on the only variable $E$. Equation \eqref{SCHII} is in
accordance with the scaling behavior of $\chi'' T^{0.66}$
experimentally established in Ref. \cite{herb3}. As it was done for
the effective mass when constructing \eqref{UN2}, we introduce the
dimensionless function
$(T^{2/3}\chi'')_{N}=T^{2/3}\chi''/(T^{2/3}\chi'')_{\rm max}$ and
the dimensionless variable $E_N=E/E_{\rm max}$, and Eq.
\eqref{SCHII} is modified to read
\begin{equation}\label{SCHIN}
(T^{2/3}\chi'')_N\simeq\frac{b_1E_N}{1+b_2E_N^2},
\end{equation}
with $b_1$ and $b_2$ are fitting parameters which are to adjust the
function on the right hand side of Eq. \eqref{SCHIN} to reach its
maximum value 1 at $E_n=1$.

\begin{figure}[!ht]
\begin{center}
\includegraphics [width=0.47\textwidth]{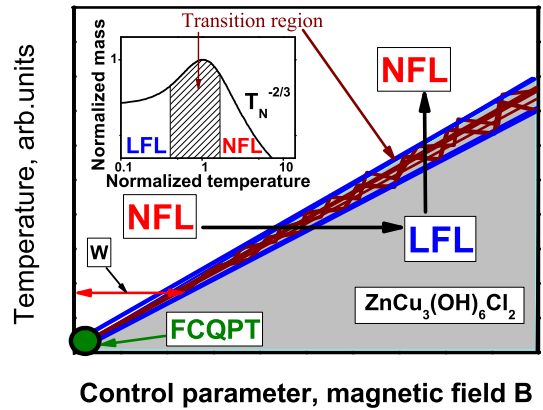}
\vspace*{-1.6cm}
\end{center}
\caption{(color online). Schematic $T-B$ phase diagram of $\rm
ZnCu_3(OH)_6Cl_2$ located on the disordered side of FCQPT. The
solid circle at the origin shown by the arrow represents FCQPT.
Vertical and horizontal arrows show LFL-NFL and NFL-LFL transitions
at fixed $B$ and $T$ respectively. The total width $W$ of the NFL
and the transition region is shown by the arrow. The inset
demonstrates the behavior of the normalized effective mass $M^*_N$
versus normalized temperature $T_N$ as given by Eq. \eqref{UN2}.
Temperatures $T_N\sim 1$ signify a transition region between the
LFL regime with almost constant effective mass and NFL one, given
by $T^{-2/3}$ dependence. The transition region, where $M^*_N$
reaches its maximum at $T/T_{\rm max}=1$, is shown by the arrows
and hatched area both in the main panel and in the
inset.}\label{fig1}
\end{figure}
Now we construct the schematic $T-B$ phase diagram of $\rm
ZnCu_3(OH)_6Cl_2$ reported in Fig. \ref{fig1}. At $T=0$ and $B=0$
the system is near FCQPT without tuning. It can also be shifted
from FCQPT by the application of magnetic field $B$. Magnetic field
$B$ and temperature $T$ play the role of the control parameters,
driving it from the NFL to LFL regions as shown by the vertical and
horizontal arrows. At fixed $B$ and increasing $T$ the system
transits along the vertical arrow from the LFL region to NFL one
crossing the transition region. On the contrary, at fixed $T$
increasing $B$ drives the system along the horizontal arrow from
the NFL region to LFL one. The inset demonstrates the universal
behavior of the normalized effective mass $M^*_N$ versus normalized
temperature $T_N$ as given by Eq. \eqref{UN2}. It follows from Eq.
\eqref{UN2} and seen from Fig. \ref{fig1} that the total width $W$
of the NFL and the transition region, shown by the arrow in Fig.
\ref{fig1}, tends to zero at diminishing $T$ and $B$ since
$W\propto T\propto B$.

\begin{figure}[! ht]
\begin{center}
\includegraphics [width=0.47\textwidth]{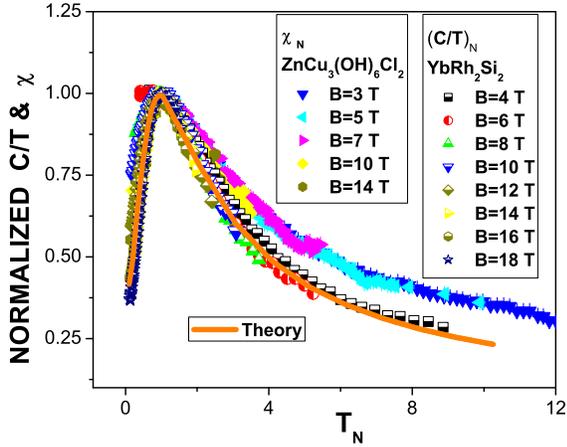}
\end{center}
\caption{(color online). The experimental data on measurements of
$\chi_N\simeq (C/T)_N\simeq M^*_N$ and our calculations of $M^*_N$
at fixed magnetic field are shown in the legends by points of
different shape and solid curve respectively. It is clearly seen
that the data collected on both $\rm ZnCu_3(OH)_6Cl_2$ \cite{herb3}
and $\rm YbRh_2Si_2$ \cite{steg1} merge into the same curve,
obeying the scaling behavior. This demonstrates that SCQSL of
herbertsmithite is close to FCQPT and behaves like HF liquid of
$\rm YbRh_2Si_2$ in magnetic fields.}\label{fig01}
\end{figure}
A few remarks are in order here. Equation \eqref{SCHII} is valid
provided that the system approaches FCQPT from the disordered side
as shown in the phase diagram \ref{fig1}. If the system is located
on the ordered side then at $B=0$ the behavior of the effective
mass as a function of $T$ is given by \cite{pr,ks}
\begin{equation}
M^*(T)\simeq a_{\tau}T^{-1}\label{MT},
\end{equation}
where $a_{\tau}$ is a constant. Upon taking into account Eq.
\eqref{MT} and acting in the same way as it was done when deriving
Eq. \eqref{SCHII}, we obtain that the imaginary part
$\chi''(T,\omega)$ is given by the equation
\begin{equation}\label{SCHIT}
T\chi''(T,\omega)\simeq\frac{a_5E}{1+a_6E^2},
\end{equation}
where $a_5$ and $a_6$ are constants, and $E=\omega/k_BT$. It is
seen from Eq. \eqref{SCHIT} that $T\chi''(T,\omega)$ depends on the
only variable $E=\omega/k_BT$. Thus, Eqs. \eqref{SCHII} and
\eqref{SCHIT} establish two types of scaling behavior of
$\chi''(\omega,T)$. Since the scaling behavior of
$\chi''(\omega,T)$ is defined by the dependence of $M^*$ on $T$,
one may expect new types of scaling especially at the transition
region shown in Fig. \ref{fig1}.

Figure \ref{fig01} reports the behavior of the normalized $\chi_N$
and specific heat $(C/T)_N$  extracted from measurements on $\rm
ZnCu_3(OH)_6Cl_2$ \cite{herb3} and $\rm YbRh_2Si_2$ \cite{steg1},
correspondingly. It follows from Fig. \ref{fig01} that in
accordance with Eq. \eqref{SCB} the behavior of $\chi_N$ coincides
with that of $(C/T)_N$ in $\rm YbRh_2Si_2$.

\begin{figure} [! ht]
\begin{center}
\includegraphics [width=0.47\textwidth]{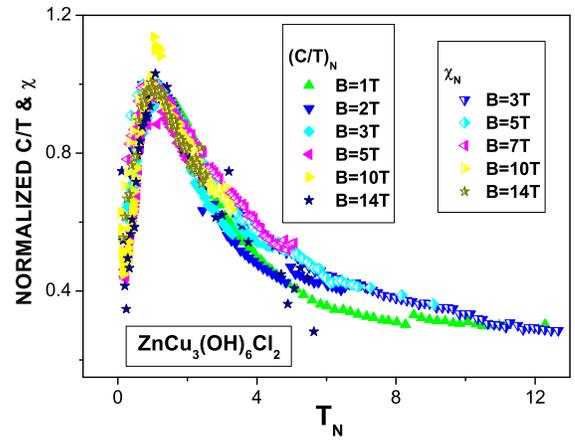}
\end{center}
\caption{(color online). The normalized susceptibility
$\chi_N\simeq M^*_N$ and the normalized specific heat
$(C/T)_N\simeq M^*_N$ of SCQSL versus normalized temperature $T_N$
as a function of the magnetic fields shown in the legends. $\chi_N$
and $(C/T)_N$ are extracted from the data of \cite{herb3} and
\cite{herb1}, respectively.}\label{fig23}
\end{figure}
In Fig. \ref{fig23} the normalized $\chi_N$ and $(C/T)_N$,
extracted from measurements on $\rm ZnCu_3(OH)_6Cl_2$
\cite{herb3,herb1}, are depicted. To extract the specific heat $C$
coming from the contribution of SCQSL from the total specific heat
$C_t(T)$ measured on $\rm ZnCu_3(OH)_6Cl_2$, we approximate
$C_t(T)$ at $T>2$ K by the function \cite{eplh}
\begin{equation}\label{CT}
C_t(T)=a_1T^3+a_2T^{1/3},
\end{equation}
where the first term proportional to $a_1$ is due to the lattice
(phonon) contribution and the second one is determined by SCQSL
when it exhibits the NFL behavior as it follows from Eq.
\eqref{MTT}, $C\propto TM^*\propto T^{1/3}$. Taking into account
that the phonon contribution is $B$-field independent, we obtain
$C(B,T)=C_t(B,T)-a_1T^3$. It is seen from Figs. \ref{fig01} and
\ref{fig23} that $(C/T)_N\simeq\chi_N$ displays the same scaling
behavior as $(C/T)_N$ measured on the HF metal $\rm YbRh_2Si_2$.
Therefore, the scaling behavior of the thermodynamic functions of
herbertsmithite is the intrinsic feature of the compound and has
nothing to do with magnetic impurities. The observed scaling
behavior of $\chi_N$ and $(C/T)_N$ in magnetic fields shown in
Figs. \ref{fig01} and \ref{fig23} rules out a possible supposition
that extra $\rm Cu$ spins outside the kagome planes considered as
weakly interacting impurities could be responsible for the
divergent behavior of the susceptibility at low temperatures.

\begin{figure} [! ht]
\begin{center}
\includegraphics [width=0.47\textwidth]{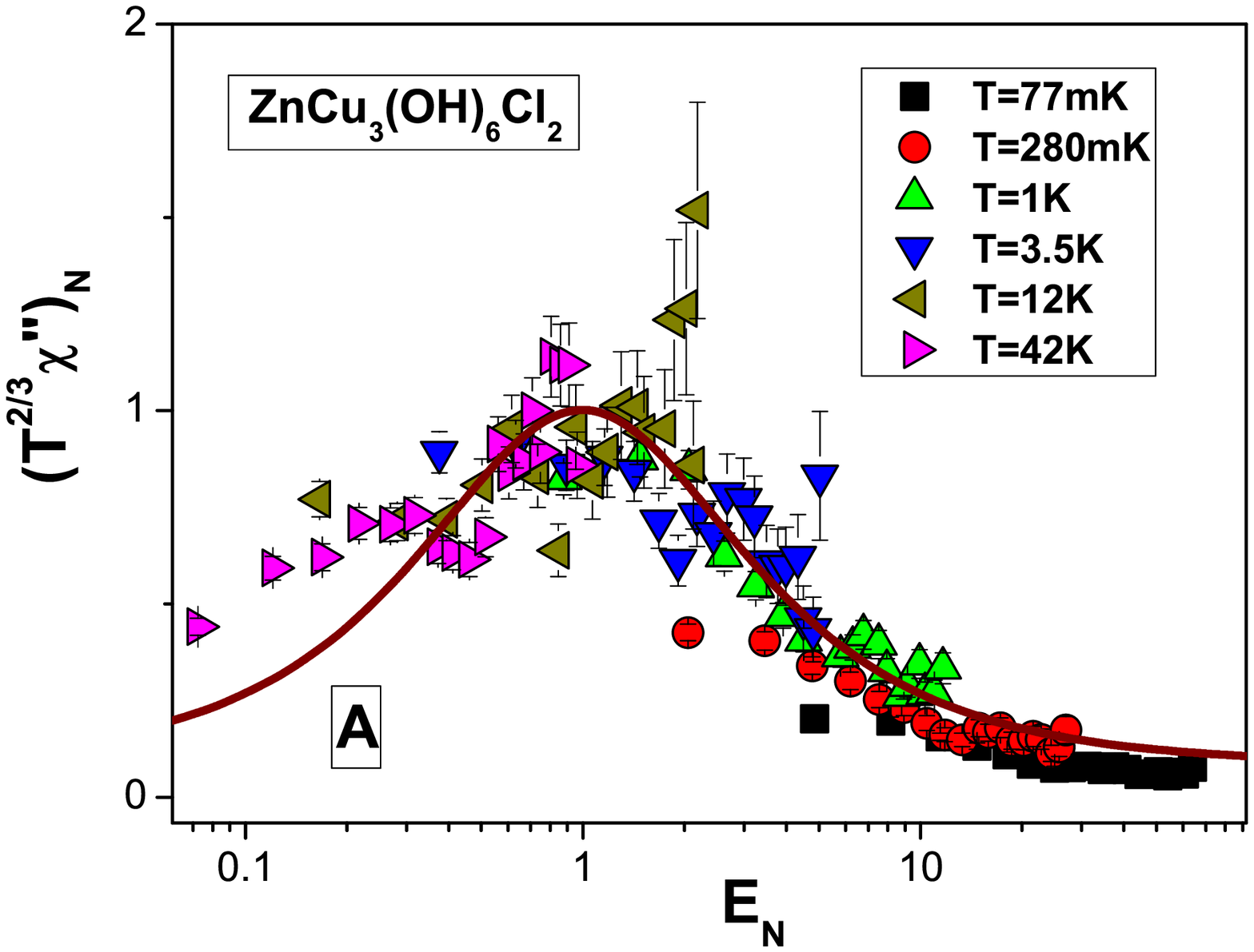}
\includegraphics [width=0.47\textwidth]{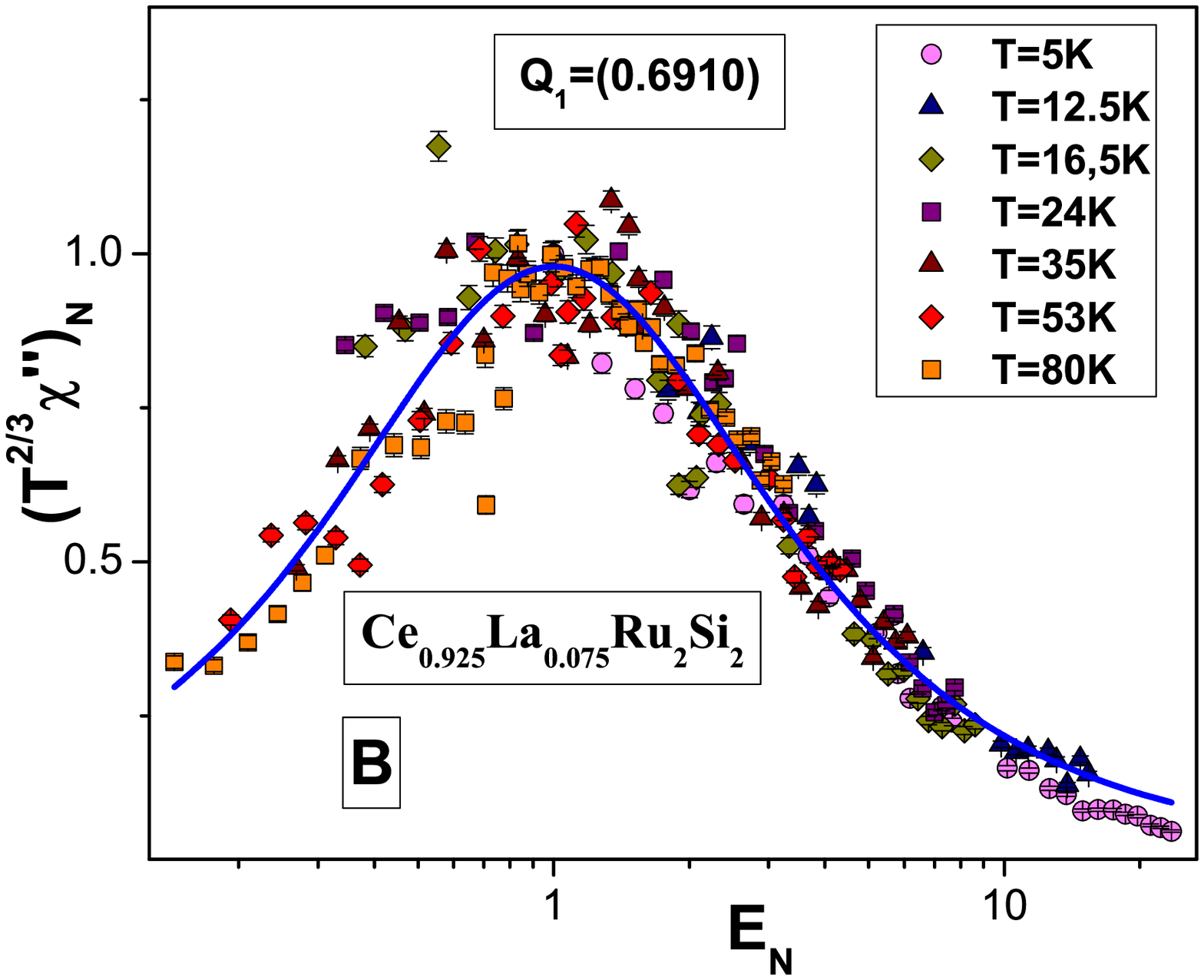}
\includegraphics [width=0.42\textwidth]{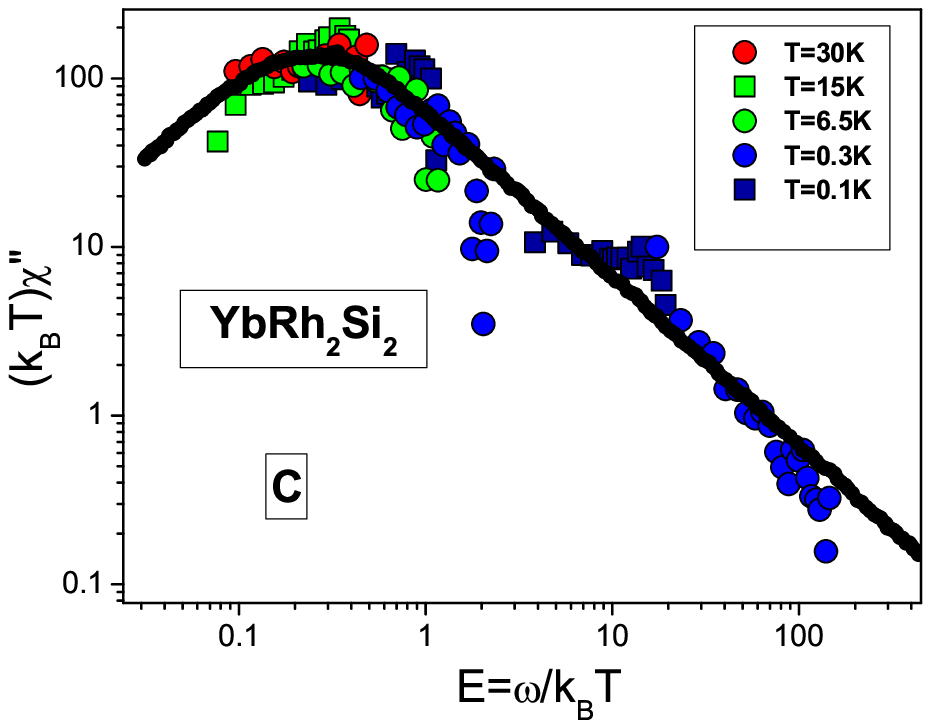}
\end{center}
\caption{(color online). Panels A and B, $(T^{2/3}\chi'')_N$
plotted against the unitless ratio $E_N=\omega/((k_BT)^{2/3}E_{\rm
max})$. The data are extracted from measurements on $\rm
ZnCu_3(OH)_6Cl_2$ \cite{herb3}, Panel A, and Panel B, on $\rm
Ce_{0.925}La_{0.075}Ru_2Si_2$ obtained at $Q_1$ \cite{prbh}. Panel
C, $T\chi''$ plotted against $E=\omega/k_BT$. The data are
extracted from measurements on $\rm YbRh_2Si_2$ \cite{stock}. The
solid curves, Panels A and B, are fits with the function given by
Eq. \eqref{SCHIN}, Panels C, with the function given by Eq.
\eqref{SCHIT}.}\label{fig2}
\end{figure}
In Fig. \ref{fig2}, consistent with Eq. \eqref{SCHIN}, the scaling
of the normalized dynamic susceptibility $(T^{2/3}\chi'')_N$
extracted from the inelastic neutron scattering spectrum of both
herbertsmithite \cite{herb3}, Panel A, and the HF metal $\rm
Ce_{0.925}La_{0.075}Ru_2Si_2$ \cite{prbh}, Panel B, is displayed.
In Fig. \ref{fig2}, Panel C, the dynamic susceptibility $(T\chi'')$
extracted from measurements of the inelastic neutron scattering
spectrum on the HF metal $\rm YbRh_2Si_2$ \cite{stock} is shown.
The data $(T\chi'')$ exhibit the scaling behavior over three
decades of the variation of both the function and the variable,
thus confirming the validity of Eq. \eqref{SCHIT}. The scaled data
obtained in measurements on such quite different strongly
correlated systems as $\rm ZnCu_3(OH)_6Cl_2$, $\rm
Ce_{0.925}La_{0.075}Ru_2Si_2$ and $\rm YbRh_2Si_2$ collapse fairly
well onto a single curve over almost three decades of the scaled
variables. It is seen that our calculations shown by the solid
curves are in good agreement with the experimental facts. Some
remarks on a role of both the disorder and the anisotropy are in
order. The anisotropy is supposed to be related to the
Dzyaloshinskii-Moriya interaction, exchange anisotropy, or
out-of-plane impurities. Measurements of the susceptibility on the
single crystal of herbertsmithite have shown that it closely
follows that measured on a powder sample \cite{herb,sl1}. At low
temperatures $T\lesssim70$ K, the single-crystal data do not show
substantial magnetic anisotropy \cite{herb,sl1}. These confirm that
the stoichiometry, disorder and anisotropy do not contribute
significantly to the results at relatively low temperatures. As we
have seen above, the scaling behavior of the thermodynamic and
relaxation properties of herbertsmithite is its intrinsic feature
and has nothing to do with the impurities \cite{eplh}. These
observations are in agreement with a general consideration of
scaling behavior of HF metals \cite{pr}.

In summary, we have considered the non-Fermi liquid behavior and
the scaling one of such strongly correlated Fermi systems as
insulator $\rm ZnCu_3(OH)_6Cl_2$ and HF metals $\rm
Ce_{0.925}La_{0.075}Ru_2Si_2$ and $\rm YbRh_2Si_2$, and shown that
these are described within the frame of the theory of FCQPT. Our
analysis of the dynamic magnetic susceptibility of strongly
correlated Fermi systems suggests that there exist at least two
types of its scaling. We calculate the thermodynamic and relaxation
properties of herbertsmithite and HF metals. The calculations are
in a good agreement with experimental data and allow us to identify
the low-temperatures behavior of $\rm ZnCu_3(OH)_6Cl_2$ determined
by SCQSL as that observed in heavy fermion metals. Thus,
herbertsmithite can be viewed as a new type of strongly correlated
electrical insulator that possesses properties of heavy-fermion
metals with one exception: it resists the flow of electric charge.

KGP acknowledges funding from the Ural Branch of the Russian
Academy of Sciences, basic research program no. 12-U-1-1010, and
the Presidium of the Russian Academy of Sciences, program
12-P1-1014.

\end{document}